\documentclass[11pt]{article}
\usepackage{moriond,epsfig}
\usepackage{color, amsmath, amssymb, array, multirow, psboxit, rotating, mathrsfs}
\usepackage{sidecap}
\usepackage{wrapfig}
\usepackage{floatflt,graphics}

\def\be{\begin{equation}}
\def\ee{\end{equation}}
\def\bea{\begin{eqnarray}}
\def\eea{\end{eqnarray}}

\begin{document}
\vspace*{3.9cm}
\title{HIGH $Q^2$ CROSS SECTIONS, ELECTROWEAK MEASUREMENTS AND PHYSICS BEYOND STANDARD MODEL AT HERA}

\author{R. PLA\v CAKYT\. E (on behalf of the H1 and ZEUS Collaborations) }
\address{DESY, Notkestr 85, 22607 Hamburg, Germany}

\maketitle\abstracts{
HERA, the only $e^{\pm}p$ collider, operated in the years 1992-2007 at centre-of-mass energies 
of 300-320 GeV in Hamburg, Germany.    
Deep inelastic neutral ($ep \rightarrow eX$) and charged 
($ep \rightarrow \nu X$) current scattering at HERA
provides the possibility to study the structure of the proton, the dynamics
of strong interactions and test quantum chromodynamics (QCD) over a huge kinematic range.
Both, neutral and charged current interactions provide complementary information on the QCD and electroweak 
(EW) parts of the Standard Model (SM).
In addition, HERA data allow for indirect searches for new phenomena originating at large scales,
however, no evidence for the new physics has been found.
The full statistics of data collected at HERA have been used for most of the measurements performed 
by H1 and ZEUS which are presented in this document.
}

\section{Introduction}
After the upgrade of HERA in the year 2000, the specific luminosity has been increased by a factor 
of about four and longitudinally polarised lepton beams have been provided to 
the collider experiments ("HERA II" period).
The average polarisation of the lepton (electron or positron) beam 
$P_e$~\footnote{The polarisation of the lepton beam 
$P_e$ is defined as $P_e = (N_R~-~N_L)~/~(N_R~+~N_L)$ where $N_R$ ($N_L$) is the number of 
right (left) handed leptons in the beam.}
achieved at HERA typically varied between 35$\%$ to 40$\%$. 
Measurements of deep inelastic scattering (DIS) with polarised leptons on protons allow the 
parton distribution functions (PDFs) of the proton to be further constrained through polarisation 
asymmetries and specific tests of the electroweak parts of the Standard Model to be 
performed~\cite{beyer}.
For example, a test of the $V$~--~$A$ structure of charged current interactions can be performed
by measuring the polarisation dependence of the charged current cross section. This tests the origin 
of absence of the right-handed weak currents. 
\\
The collected luminosity at HERA of high energy $ep$ interactions 
(about 1 $fb^{-1}$ by H1 and ZEUS experiments together) gives access to rare processes
with cross sections of the order of 0.1 pb, providing a testing ground for the Standard Model
complementary to $e^+e^-$ and $p \overline p$ scattering. 
New phenomena at large scales (beyond the maximal available center of mass energy) may be 
detected indirectly as deviations from the SM predictions. 
For example the model independent general signature based analysis is based on a search 
differences between data and SM expectations in various event topologies.
\\
Most of results obtained by H1 and ZEUS collaborations at the high negative four-momentum
transfer squared $Q^2$ as well as of searches for new physics have been combined
in order to improve precision and sensitivity.
A part of these results are presented in this document.

\section{High $Q^2$ Cross Sections and Electroweak Physics}
The deep inelastic neutral current (NC) scattering cross section is defined as:
\\ [-0.9cm]
\begin{center}
  \begin{equation}
     \frac{d^2\sigma_{NC}^{e^{\pm} p}}{dxdQ^2}=\frac{2\pi\alpha^2}{xQ^4} 
      \Big [ Y_{+} \tilde F_2^{\pm} \mp Y_{-}x \tilde F_3^{\pm} - y^2 \tilde F_L^{\pm} \Big ].
  \label{eq:NCsigma}
  \end{equation}
\end{center}
Here, $Y_{\pm} = 1 \pm (1-y)^2$ is the helicity factor, 
$\tilde F_2^{\pm}$, $\tilde F_3^{\pm}$ and $\tilde F_L^{\pm}$ are generalised structure 
functions:
\\ [-0.9cm]
\begin{center}
 \begin{equation}
   \begin{array}{rll}
   \vspace{0.2cm}    
     \tilde F_2^{\pm} = F_2 + k\big ( - v_e \mp P_e a_e \big ) F_2^{\gamma Z} 
               + k^2 \Big ( v_e^2 + a_e^2 \pm 2 P_e v_e a_e \big ) F_2^Z   
     \\
     x\tilde F_3^{\pm} = {F_2 +} k \big ( - a_e \mp P_e v_e \big ) xF_3^{\gamma Z} 
               + k^2 \big ( 2v_e a_e \pm P_e (v_e^2 + a_e^2) \big ) xF_3^Z .
   \end{array}
 \label{eq:NC_F2}
 \end{equation}
\end{center}
Pure photon exchange is described by $F_2$, pure $Z$ exchange by $F_2^Z$ and 
$xF_3^Z$, and $\gamma Z$ interference by $F_2^{\gamma Z}$ and $xF_3^{\gamma Z}$. 
$v_e$ is the weak vector coupling and $a_e$ the weak axial-vector coupling of the electron 
to the $Z$. The quantity $k$ is defined via the Weinberg angle $\theta_w$, the four-momentum
transfer squared $Q^2$ and mass of the Z boson $M_Z$: $k = \frac{1}{4 \sin^2 \theta_w
\cos^2 \theta_w} \frac{Q^2}{Q^2 + M_Z^2}$.   
As can be seen from Eq.~\ref{eq:NCsigma}, the polarised lepton beams modify the neutral current cross 
sections mostly via the $\gamma Z$ interference and Z terms. 
\\
The charged current (CC) cross section is defined as:
\\ [-1.0cm]
\begin{center}
  \begin{equation}
     \frac{d^2\sigma_{CC}^{e^{\pm} p}}{dxdQ^2} = \big( 1 \pm P_e \big)
            \frac{G_F^2}{2\pi x} \bigg( \frac{M_W^2}{Q^2+M_W^2} \bigg)^2 \tilde \sigma_{CC}^{e^{\pm} p},
 \label{eq:polCCsigma}
  \end{equation}
\end{center}
where the reduced charged current cross section $\tilde \sigma_{CC}^{e^{\pm} p}$
is related to the quark densities in $e^{\pm}p$ scattering~via \\ [-1.0cm]
\begin{center}
  \begin{equation}
    \begin{array}{rll}
   \vspace{0.2cm}    
   e^{+}:  & &  \tilde \sigma_{CC}^{e^{+} p} = x[\overline u +\overline c] + (1-y)^2 x[ d+s ]  \\
   e^{-}:  & &  \tilde \sigma_{CC}^{e^{-} p} = x[u +c] + (1-y)^2 x[\overline d +\overline s ].
    \end{array}
    \label{eq:CCstrucFuncProbQuark}
  \end{equation}
\end{center}
From Eq.~\ref{eq:polCCsigma} can be seen that in the Standard Model the charged current cross 
section has a linear 
dependence on $P_e$. Therefore the charged current cross section is zero for fully right (left)
handed electron (positron) beam providing a consistency test of the electroweak sector of the Standard 
Model.
\begin{figure}[h]
 \begin{minipage}[b]{0.48\linewidth}
   \centering
     \includegraphics[width=17.5pc]{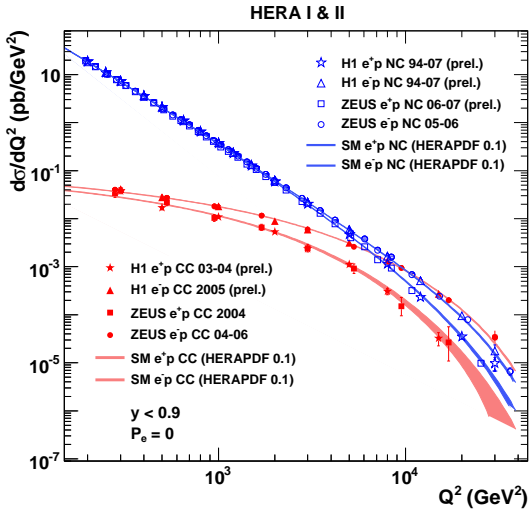}
 \end{minipage}
\hspace{0.1cm} 
 \begin{minipage}[t]{0.47\linewidth}
      \vspace{-6.8cm}
       \centering
       \epsfig{file=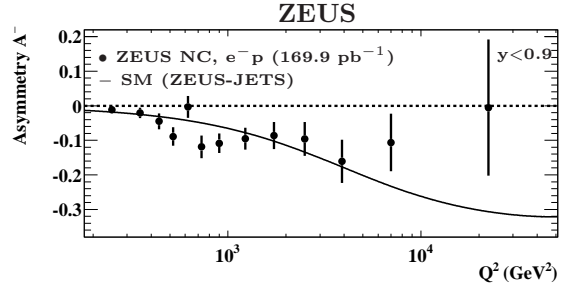,bburx=160,bbury=130,bbllx=430,bblly=700,clip=,height=7.9cm, angle=90} 
       \put (-127,100) { \small \textbf{ZEUS} }
       \put (-44,  86) { \tiny \textbf{y$<$0.9} }
       \put (-190, 82) {\textcolor{white}{ \rule[0.3mm]{6mm}{4mm} } }
       \put (-195, 85) { \tiny \textbf{$\bullet$ ZEUS NC, e$^-$p (169.9 pb$^{-1}$)}}
       \put (-195, 76) { \tiny \textbf{$-$ SM (ZEUS-JETS)}}
   \centering
   \caption{\it The $Q^2$ dependences of the neutral and charged current cross sections 
             $d\sigma / dQ^2$ compared to SM predictions determined from the 
             HERAPDF 0.1 fit (left). 
             The polarisation asymmetry $A^{-}$ as a function of $Q^2$ (measured by ZEUS) 
             compared to SM predictions evaluated using ZEUS-JETS PDFs (right). }    
 \label{dsdq2_ccnc_and_zeus_nc}
 \end{minipage}
\end{figure}
\\
\\
Unpolarised neutral and charged current cross sections measured with the H1 and ZEUS detectors at 
HERA as function of $Q^2$ are shown in Figure \ref{dsdq2_ccnc_and_zeus_nc} (left). 
NC and CC cross sections become about equal in magnitude at 
$Q^2 \gtrsim 10^4 \ $GeV$^2$. 
This follows from the propagator term $Q^2$ dependence which is different for NC and CC 
up to $Q^2 \lesssim M_{Z(W)}^2$ (see Eq.~\ref{eq:NCsigma} and Eq.~\ref{eq:polCCsigma}).
\\
An access to electroweak effects is also enabled by measuring the charge dependent polarisation 
asymmetry in neutral currents. The polarisation asymmetry measures a product of vector and
axial-vector couplings and thus is a direct measure of parity violation.
Neglecting the Z term in Eq.~\ref{eq:NC_F2} and taking into account that at leading order 
$F_2^{\gamma Z} = x \sum 2 e_q v_q (q+ \overline q)$,   
then to a good approximation the polarisation asymmetry $A^{\pm}$ measures the structure function
ratio: \\ [-0.8cm]
\begin{center}
  \begin{equation}
   \nonumber
   A^{\pm} = \frac{2}{P_R - P_L} 
        \frac {\sigma^{\pm}(P_R) - \sigma^{\pm}(P_L)}{\sigma^{\pm}(P_R) + \sigma^{\pm}(P_L)} 
        \simeq \mp k a_e \frac {F_2^{\gamma Z}}{F_2},
  \end{equation}
\end{center}
which is proportional to $a_e v_q$ combinations and thus directly measures the parity violation.
In Figure \ref{dsdq2_ccnc_and_zeus_nc} (right) $A^{+}$ measured by the ZEUS 
collaboration is presented. 
\vspace{-0.1cm}
\\
\begin{floatingfigure}[r]{0.545\textwidth}
  \vspace{-0.6cm}
  \begin{center}
   \includegraphics[width=16.7pc]{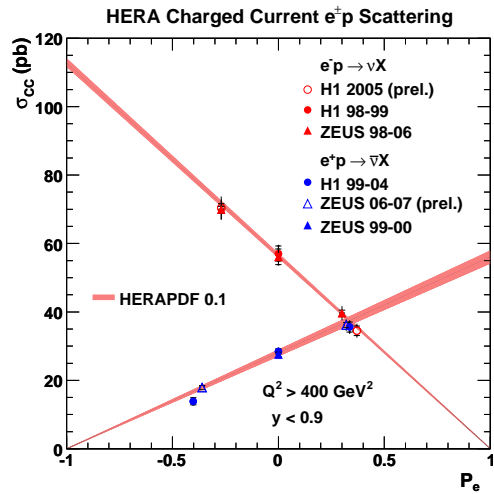}
  \end{center}
  \vspace{-0.55cm}
   \caption{\it The dependence of the charged current cross section on the lepton beam polarisation $P_e$
             at HERA.
             The data are compared to SM predictions based on HERA\-PDF~0.1 parametrisation.}    
 \label{cc_tot_and_zeus_ew}
\end{floatingfigure}
\vspace{0.1cm}
The polarisation dependence of the charged current cross sections measured by the H1 and ZEUS 
collaborations is presented in Figure \ref{cc_tot_and_zeus_ew}. This found to be 
consistent with the absence of right handed charged currents as predicted by the SM.
\\
Using the NC and CC cross sections a combined electroweak and QCD analysis was performed by the H1 and 
ZEUS collaborations in order to determine electroweak parameters~\cite{ewqcd}. In this fit a
similar or even better precision was achieved compared to fit performed by 
Tevatron and LEP experiments. 
\\
\vspace{0.3cm}

\section{Physics Beyond Standard Model}
\begin{figure}[h]
 \begin{minipage}[b]{0.3\linewidth}
   \centering
   \includegraphics[width=12.5pc]{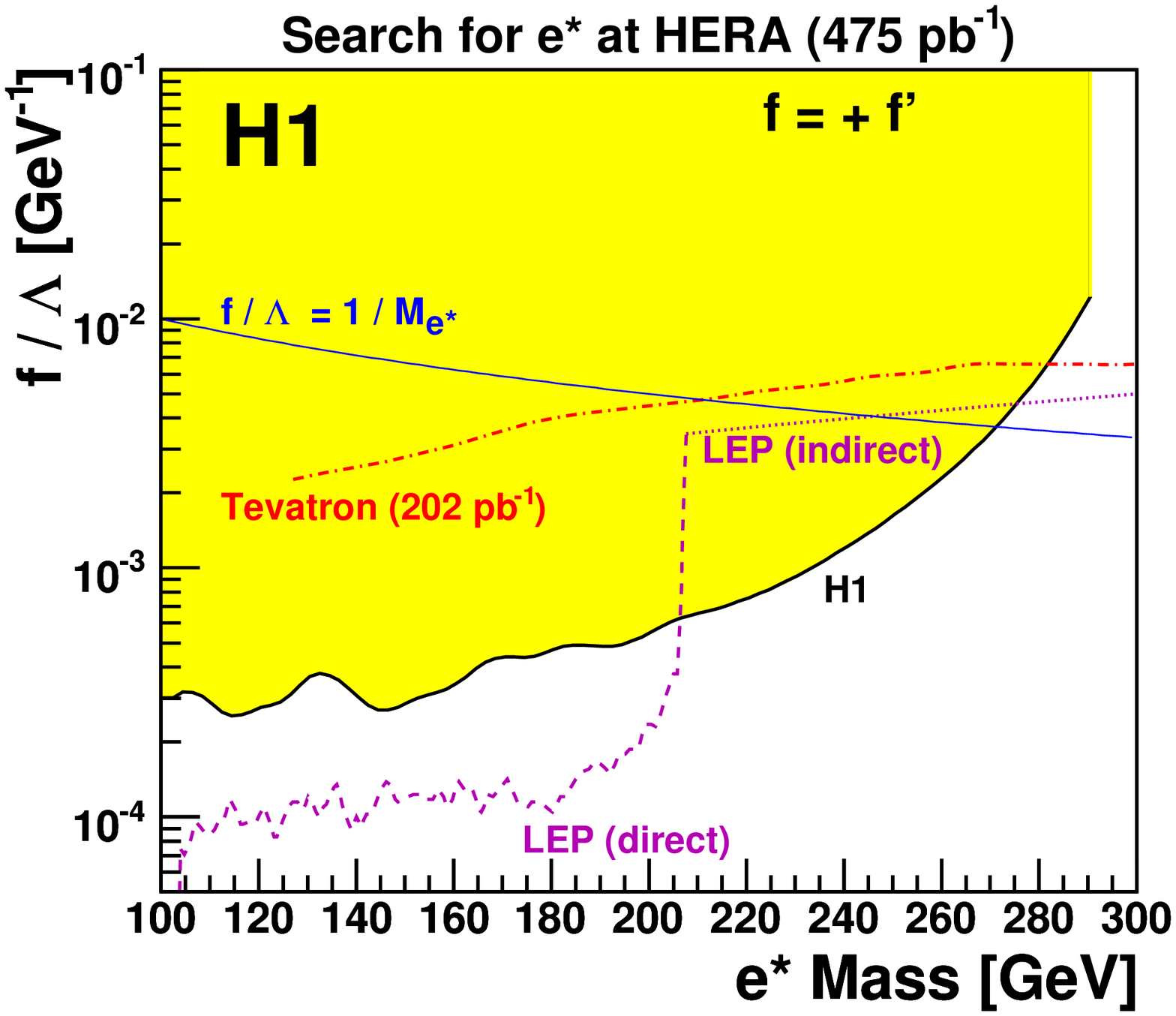}
 \end{minipage}
\hspace{0.53cm} 
 \begin{minipage}[b]{0.3\linewidth}
   \centering
   \includegraphics[width=12.5pc]{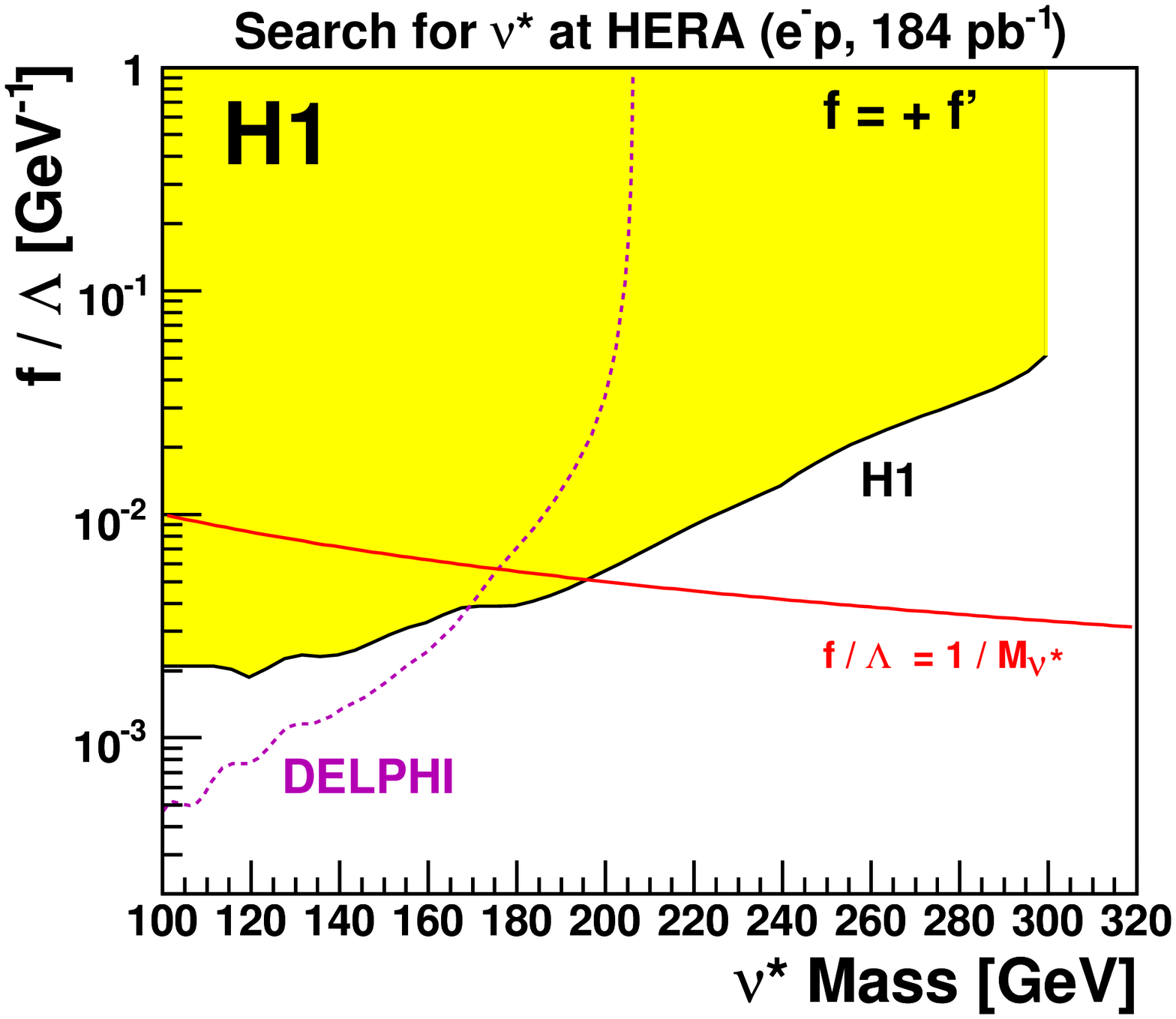}
 \end{minipage}
\hspace{0.53cm} 
 \begin{minipage}[b]{0.3\linewidth}
   \centering
   \includegraphics[width=12.5pc]{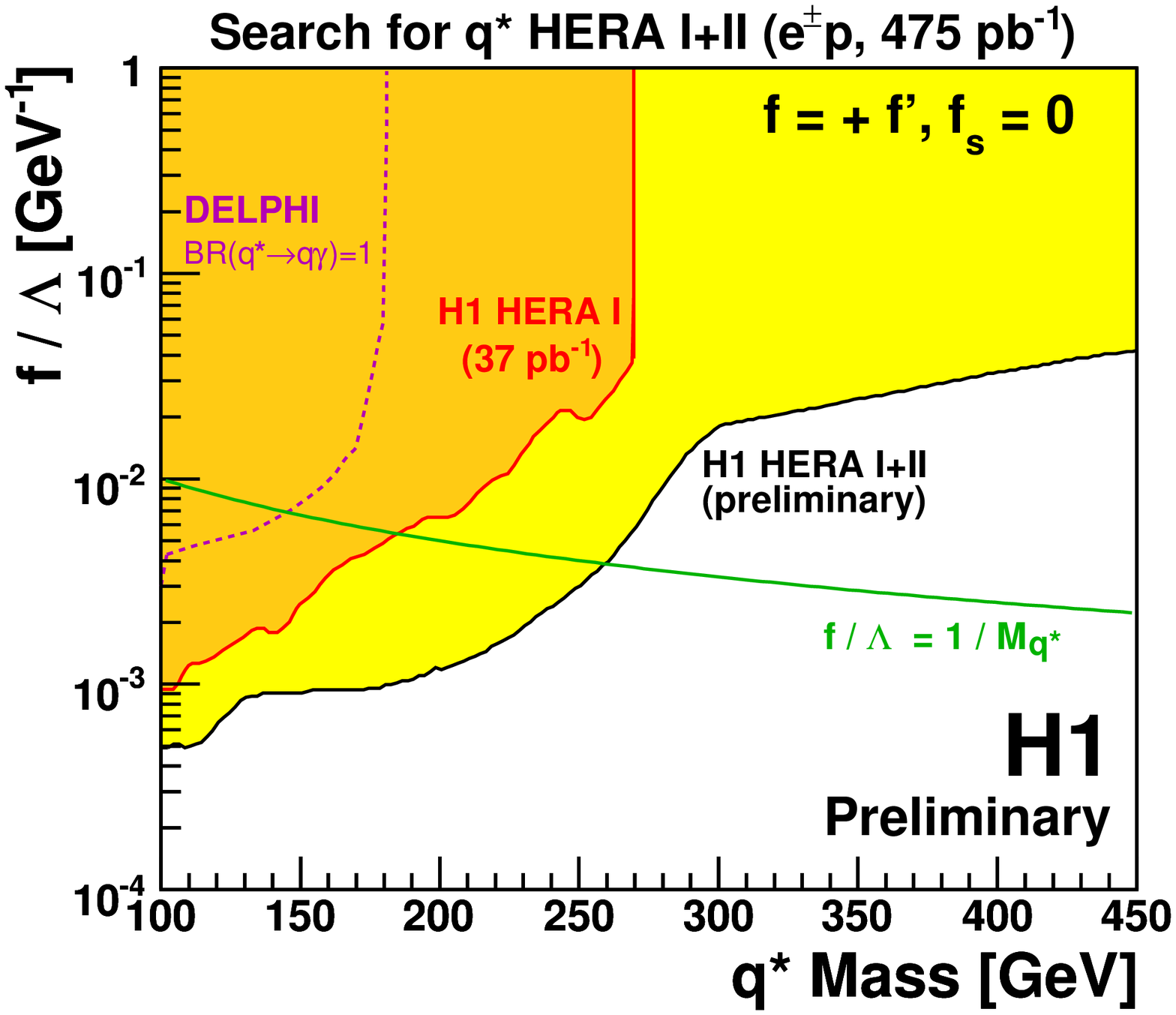}
 \end{minipage}
  \vspace{-0.7cm}
   \caption{\it Exclusion limits at 95$\%$ CL on the coupling $f/ \Lambda$ as a function
            of the mass of the excited fermion for $f=+f'$ ($e^*$ and $\nu^*$) and 
            $f=+f', f_s=0$ ($q^*$) (for symbol explanation see text). 
            The exclusion domain represented by the shaded area is compared to
            the corresponding exclusion limits obtained at LEP and Tevatron. }    
 \label{excited_fermion_limits}
\end{figure}

A search for excited states of leptons and quarks which are predicted by models assuming composite
quarks and leptons has been performed at HERA~\cite{excQ,excE,excN}.
Generally interactions between excited and ordinary fermions may be mediated by gauge bosons
and described by an effective Lagrangian:
\\ [-1.0cm]
\begin{center}
 \begin{equation}
    \mathcal{L}_{GM} = \frac{1}{2 \Lambda} \bar F^*_R \sigma^{\mu \nu}
        \Big [ g f \frac{\tau^a}{2} W^a_{\mu \nu} + g' f' 
        \frac{Y}{2} B_{\mu \nu} + g_s f_s \frac{\lambda^a}{2} G^a_{\mu \nu} 
        \Big ] F_L + h.c. ,
 \end{equation}
\end{center}
where, $\sigma^{\mu \nu}$ is the covariant bilinear tensor, $W^a_{\mu \nu}, B_{\mu \nu}$ and 
$G^a_{\mu \nu}$ are the field-strength tensors of the $SU(2), U(1)$ and $SU(3)_{C}$ gauge fields,
$\tau^a, Y$ and $\lambda^a$ are the Pauli matrices, the weak hypercharge operator and the 
Gell-Mann matrices, respectively. The standard electroweak and strong gauge couplings are denoted
as $g, g'$ and $g_s$. The compositeness scale $\Lambda$ reflects the range of the new confinement 
force, coupling parameters $f, f'$ and $f_s$ associate the three gauge groups. \\
In the recent analysis performed in the H1 collaboration using full HERA statistics 
the excited lepton and quark decay channels (e.g. $q^* \rightarrow q \gamma, 
q^* \rightarrow qZ, q^* \rightarrow qW$ for quark) with subsequent hadronic and leptonic decays of Z/W
bosons were considered. No indication of a signal was found.
An upper limit on the coupling $f/ \Lambda$ as a function of the excited fermion mass was established
and is presented in Figure~\ref{excited_fermion_limits} (the specific relations between couplings 
are noted in corresponding figures).
The results of searches for excited fermions at HERA are complementary or have better sensitivity 
compare to similar measurements performed at LEP and Tevatron.
\begin{floatingfigure}[r]{0.49\textwidth}
  \vspace{-0.3cm}
  \begin{center}
 \begin{minipage}[t]{0.24\linewidth}
  \hspace{-1.8cm}
  \begin{rotate}{270}
   \includegraphics[width=16pc]{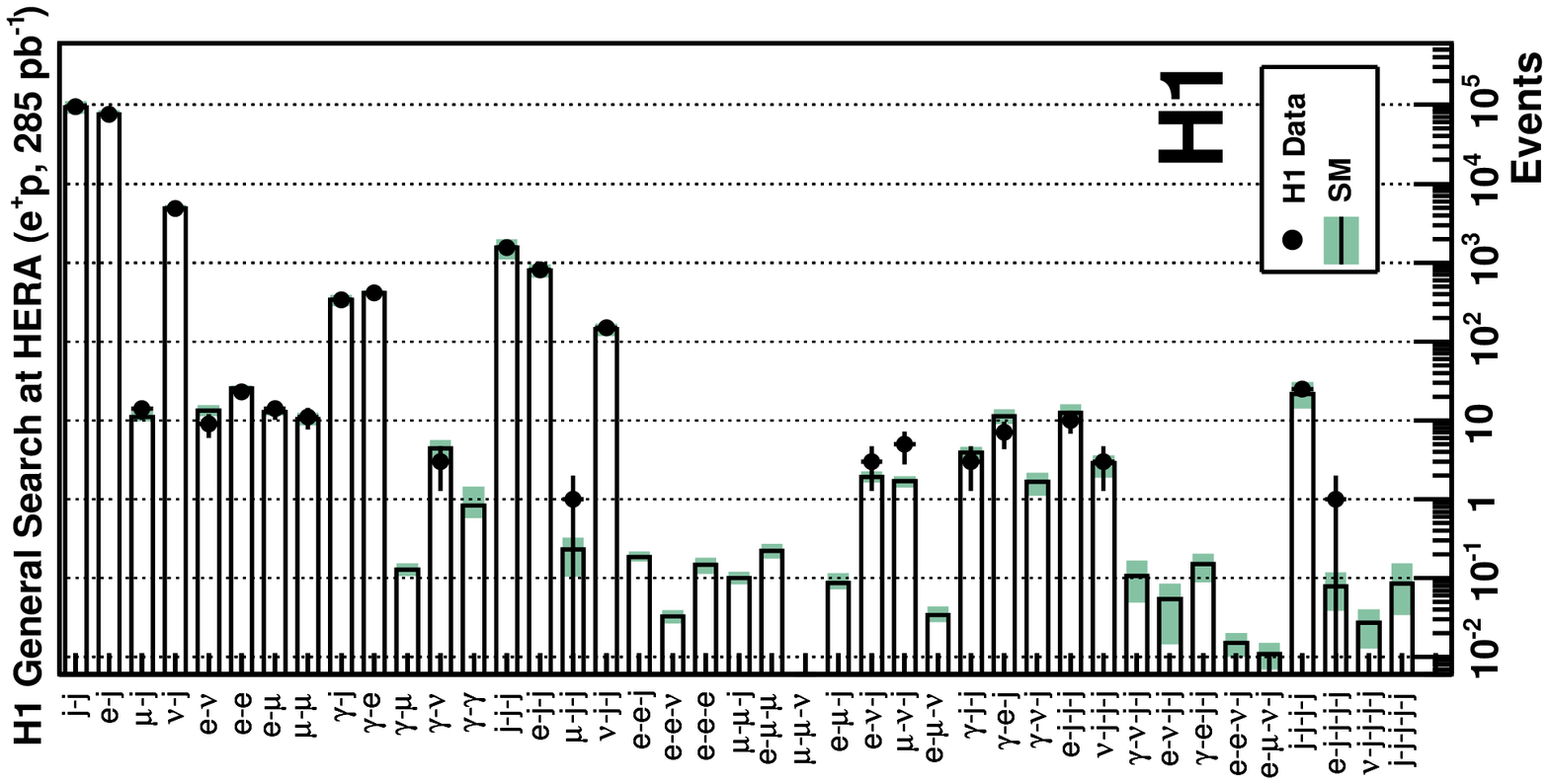}
  \end{rotate}
 \end{minipage}
\hspace{0.1cm} 
 \begin{minipage}[t]{0.24\linewidth}
  \vspace{-0.49cm}
  \hspace{1.9cm}
  \begin{rotate}{270}
   \includegraphics[width=16pc]{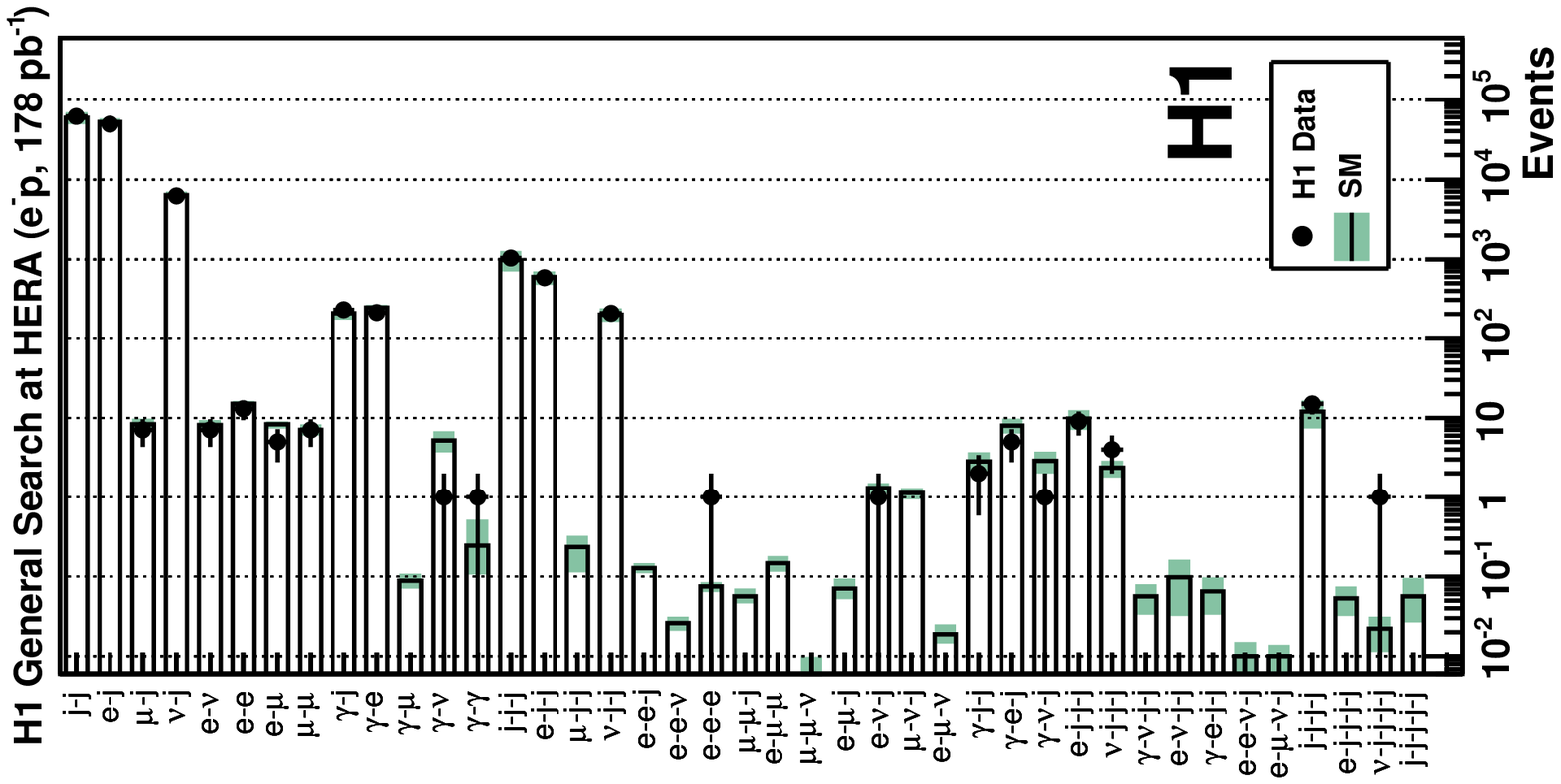}
  \end{rotate}
 \end{minipage}
  \end{center}
  \vspace{5.9cm}
   \caption{\it The event yields for data and SM expectations of 27 event classes analysed 
           in the generic search for new phenomena at HERA for $e^{+}p$ (left) and $e^{-}p$ (right)
           collisions. }    
 \label{general_searches}
\end{floatingfigure}
\vspace{0.1cm}
A generic analysis searching for deviations between DIS data and SM
expectation in all high transverse momenta $P_T$ event topologies has been recently published 
by H1 collaboration~\cite{genAnal}.
An advantage of such a generic analysis is that it does not rely on any a priori definition of 
expected signature of a specific model, i.e. is model independent.
The search has been performed for all event topologies involving isolated electrons, photons, muons,
neutrinos and jets with transverse momenta above 20 GeV. 
In each case deviations from the SM were searched for in the invariant mass and sum of transverse 
momenta distributions using a dedicated algorithm. 
A good agreement with the SM expectation was observed in the analysis which
demonstrates the very good understanding of high $P_T$ phenomena achieved at HERA. 
The event yields for analysed event classes comparing data and SM expectations are 
presented in Figure~\ref{general_searches} separately for $e^{+}p$ and $e^{-}p$ collisions. 
Other HERA results like e.g. multi-lepton analysis can be found in the physics result web-sites of 
H1 and ZEUS collaborations.
\\ [-0.7cm]

\end{document}